# Conflicting History in Foundation Model Return Forecasts

**Levent Bulut**
Department of Data Analytics and Statistics
The Anuradha and Vikas Sinha College of Artificial Intelligence
and Advanced Analytics, University of North Texas
levent.bulut@unt.edu

**Abstract**

Older observations can help a forecaster learn a predictive relationship while becoming misleading when that relationship changes. We study this tension using paired return histories that differ in older observations but share recent returns and the conditional distribution of the next return. Across five foundation-model forecasting procedures and three simulated processes, each evaluated on 1,024 independent replications, conflicting history reduces the probability of improving on zero in all 30 higher-signal comparisons. This loss of usefulness occurs despite successful signal recovery under stable conditions. A separate exploratory comparison on 56 assets finds that shortening inputs from 256 to 64 returns increases mean absolute error for every model by 1.77–4.58 basis points; neither length improves observed overall error over zero. The evidence distinguishes the ability to extract a signal from the ability to withstand conflicting history, while showing why context sensitivity alone does not justify discarding older data.

## 1 Introduction

How much history should a return forecaster use? Older observations may help reveal a weak predictive relationship, but they may also point to a relationship that no longer holds. This tension makes performance after a change difficult to interpret. A forecast may deteriorate because the current signal has weakened, because the forecaster continues to rely on older observations, or because both have occurred. Forecast accuracy alone does not distinguish these explanations.

This paper isolates the role of conflicting history. We construct pairs of return histories that share the same recent observations and the same conditional distribution of the next return, while differing in older observations. One history follows the current relationship throughout. The other contains a change from the opposite relationship. Any difference in conditional forecast risk within the pair therefore reflects the supplied older history, rather than a difference in current predictability.

The distinction matters because learning a signal and remaining useful after a change are separate forecasting capabilities. A model that predicts zero regardless of its input is perfectly insensitive to conflicting history, yet extracts no signal. Conversely, a model may learn a stable relationship well and continue to apply it after it has changed. A useful evaluation must measure both signal recovery and sensitivity to the past.

We examine five foundation-model forecasting procedures on three simulated return processes. The construction uses symmetric shocks and a sign transformation to reverse dependence while preserving volatility and the current target distribution. Positive dependence means that the conditional median continues the preceding return; negative dependence means that it reverses it. Conflicting observations are valid returns generated under the opposite relationship, rather than corrupted or artificially mislabeled data. Because the conditional law is known, forecasts can be evaluated by expected absolute error without adding noise from the next realized shock.

The results reveal a systematic gap between stable signal recovery and performance with conflicting history. At the higher of two benchmark signal levels, each model improves on zero on a majority of stable paths. Introducing conflicting older observations lowers that probability in every combination of model, reversal direction and generating process. Responses also differ by the direction of the reversal, and frequent improvements do not necessarily translate into favorable average losses. These findings motivate evaluating the frequency, magnitude and historical sensitivity of forecast gains together.

They do not, however, imply that shorter histories are generally better. We test that implication separately on an exploratory panel of 56 assets and 434 forecast origins. Holding the asset, origin, target and model fixed, we compare inputs containing 64 and 256 returns. Shortening increases mean absolute error for all five procedures, by 1.77–4.58 basis points, and both lengths have higher observed overall error than predicting zero. The historical comparison thus places an important limit on the practical interpretation of the simulation: identifying sensitivity to conflicting history does not establish a successful rule for removing observations.

The paper contributes an evaluation design that holds current predictive information fixed while varying the older history, together with evidence on how five forecasting procedures respond to that intervention. The historical comparison addresses a related practical question without assuming that the removed observations belong to an obsolete regime. Keeping these questions distinct allows the controlled result and the unfavorable historical result to inform one another.

### 1.1 Relation to the literature

Recent financial foundation-model studies examine return prediction (Noguer I Alonso & Franklin, 2026), financial pretraining and fine-tuning (Rahimikia et al., 2025), and the consequences of training vintage (Chen et al., 2026). These studies ask whether pretrained models can deliver useful forecasts and how their training affects evaluation. We focus on another source of variation: the history supplied to a fixed forecasting procedure. Matching the current target law allows us to separate sensitivity to that history from changes in the signal available to the model.

A related literature develops methods for adapting forecasts as conditions change. TAFAS updates calibration modules as new observations arrive (Kim et al., 2025), while RefineBridge learns a separate model to refine foundation-model forecasts (Bolton et al., 2025). Credal-TTA manages context to reduce the influence of obsolete patterns after regime changes (Du & Tang, 2026). The question raised by our design is whether adaptation preserves the ability to use a stable signal while limiting the influence of conflicting observations. We evaluate this distinction without comparing those adaptation methods.

Our approach also relates to controlled studies of the information used by forecasting models. Tire et al. (2026) examine synthetic motifs and retrieved examples, and Won (2026) studies responses to counterfactual external inputs in forced-system identification. Here the intervention concerns older endogenous returns. Recent observations and the next-return distribution are matched exactly, which makes it possible to attribute a difference in conditional risk to the older portion of the input within the specified experiment.

## 2 An experiment that holds current predictability fixed

### 2.1 Return processes

The simulated returns combine a predictable conditional location with changing volatility:

$$
\begin{aligned}
x_t &= q(x_{t-1}) + e_t, \quad e_t = \sqrt{h_t}\, z_t, \\
z_t &= \sqrt{(\nu-2)/\nu}\, T_t, \quad T_t \sim t_\nu, \\
h_t &= \omega + \alpha e_{t-1}^2 + \beta h_{t-1}, \quad \omega = (1-\alpha-\beta)v.
\end{aligned} \tag{1}
$$

The innovations are independent Student-t draws standardized to unit variance. The parameter $v$ is the target variance of the shocks $e_t$. In the linear process, $q(x) = ax$, with $a = 0.12$ or $0.40$. We refer to these as the lower and higher benchmark signals. They are experimental settings, rather than estimates of market predictability. The nonlinear process replaces the linear response with a saturating function, $q(x) = a\kappa \tanh(x/\kappa)$. Its amplitudes are calibrated on separate samples to approximately match the linear process's mean available improvement over zero.

A third process retains the linear signal coefficients and calibrates the shock distribution and volatility dynamics to SPY returns from 2024. The fitted Student-t parameter controls the frequency of unusually large shocks; the volatility parameters determine how their effects persist. This calibration brings financial features into the noise surrounding the signal, without fitting the strength of the predictive relationship to market data. Each process has its own fresh evaluation sample; the nonlinear and financial designs were fixed after inspecting the preceding evidence.

## 2.2 Paired histories

We obtain different dependence patterns from a common simulated path by setting $r_t = s_t x_t$, where the deterministic signs $s_t$ take values −1 or +1. Since $q$ is odd,

$$\text{median}(r_t \mid \text{past}) = s_t s_{t-1} q(r_{t-1}). \tag{2}$$

Constant signs generate positive dependence, while alternating signs generate negative dependence. Symmetric innovations preserve the shock law under this transformation, and the squared shocks preserve the volatility recursion. Choosing the sign sequences to coincide from a common anchor gives two histories with identical recent returns and an identical conditional next-return distribution. Before the anchor, their dependence can differ.

Figure 1 summarizes the paired design, and Figure 2 illustrates it with one saved pair of simulated histories.

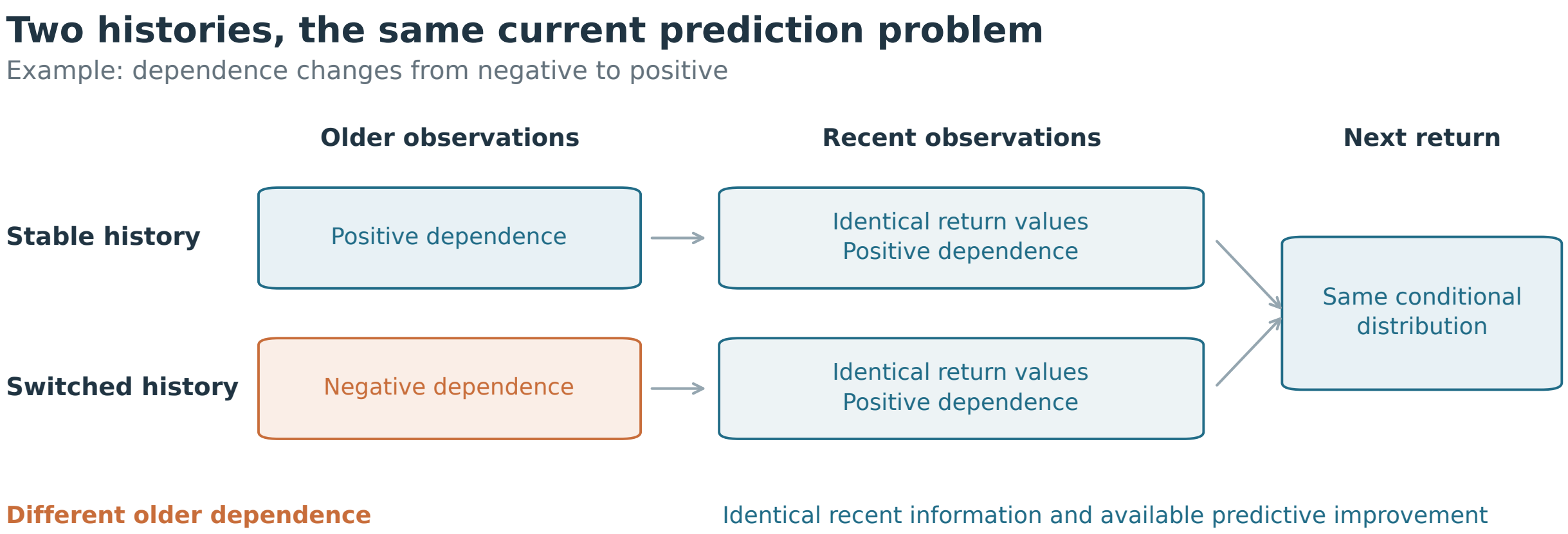

**Figure 1.** The paired comparison for a change from negative to positive dependence. The stable history follows positive dependence throughout; the switched history contains older observations generated under negative dependence. The recent suffix and conditional next-return distribution are identical. The opposite reversal is constructed in the same way. Blocks represent portions of a history, not individual returns.

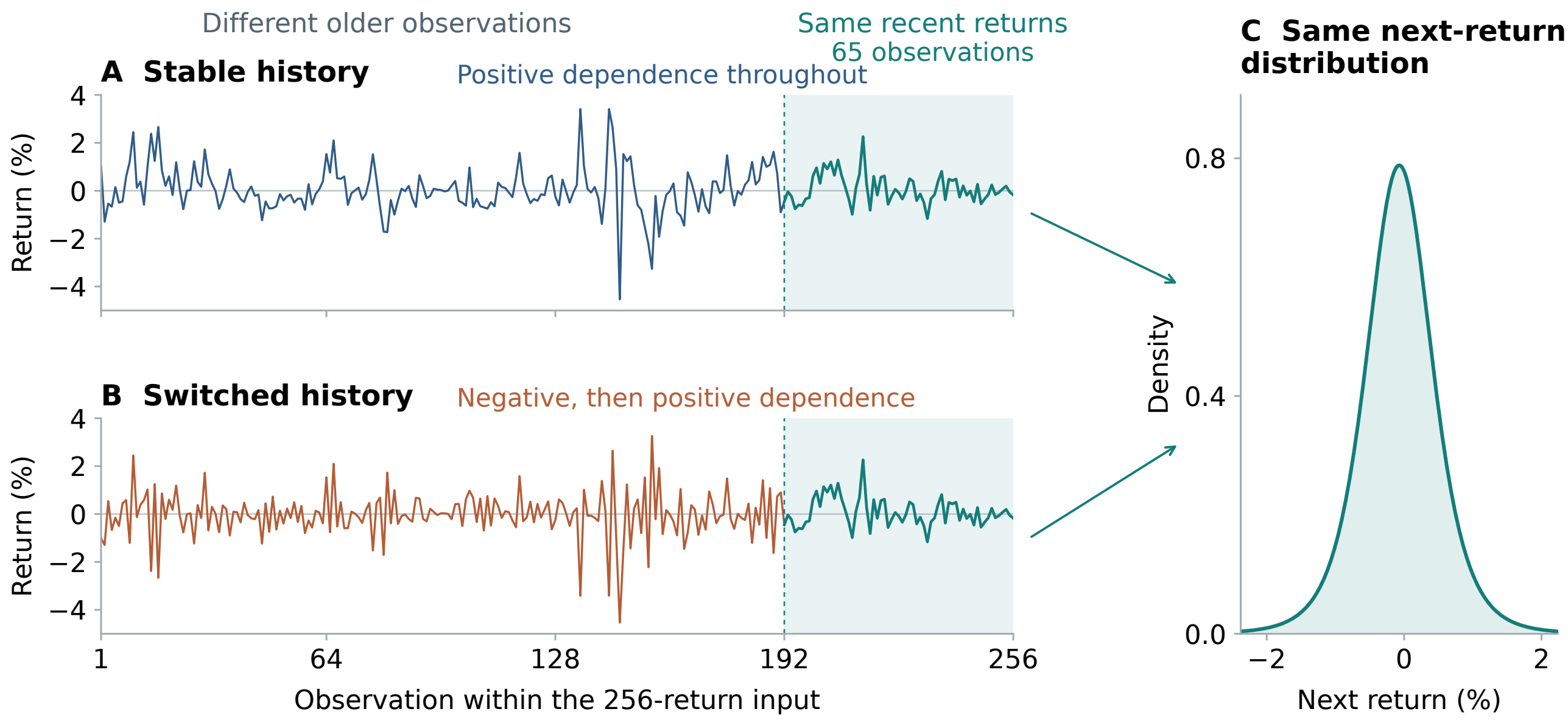

**Figure 2.** A time-series example of the paired design. Panels A and B show one saved pair of simulated 256-return inputs from the financially calibrated background, under the strong-signal setting and at offset 64. The stable history has positive dependence throughout; the switched history changes from negative to positive dependence. The shaded portions contain the same 65 returns, including the common anchor at observation 192. Panel C shows their shared conditional distribution for the next return, rather than a model forecast. Returns are expressed in percent. The opposite reversal is constructed in the same way.

In Figure 2, reading the two histories from left to right makes the comparison concrete. The older portions differ, but the teal portions match observation for observation. Both histories therefore end with the same recent data and face the same distribution of possible next returns. Comparing forecasts for the pair isolates how the older observations affect the forecasting procedure.

Every forecast uses a 256-return input. We evaluate the histories at offsets 0, 16, 64, 128 and 256 from the change, and average offsets 16, 64 and 128 for the primary comparison. The matched suffixes at those three offsets contain 17, 65 and 129 returns, including the common anchor immediately before the change. As the window moves forward, older observations leave it. The two inputs become identical at offset 255, so the saved offset-256 comparison provides an endpoint at which their histories coincide. Positive and negative throughout refer to dependence, not to rising and falling markets.

### 2.3 Forecast gains and uncertainty

The five procedures are TabPFN v3 (Prior Labs Team, 2026), Chronos-2 (Ansari et al., 2025), Moirai 2 (Salesforce, n.d.), TimesFM 3 (Google Research, n.d.) and Toto 2 22M (Datadog, n.d.), evaluated with fixed checkpoints and forecast conventions. TabPFN uses eight lag features; the other procedures use their sequence representations. We compare them with zero, the historical median, full-context AR(1) and recent-history AR(1). The recent AR model fits 64 lag pairs from 65 returns. The nonlinear experiment also includes autoregressive references using the known nonlinear feature. These references help distinguish sensitivity specific to a model from difficulty shared with a simple estimator.

Let $R(f;\mu,\sigma)$ denote the conditional expected absolute error of forecast $f$, where μ and σ are the conditional median and standard deviation. The conditional median minimizes this loss. Thus $R(0) - R(\mu)$ measures the improvement available over predicting zero when the current law is known, without knowledge of the next shock. We calculate $R$ analytically under the Student-t distribution. Within each history pair, the benchmark and the conditional law are the same.

For a given simulated path, the primary gains and history penalty are

$$\begin{aligned} G^{\mathrm{st}} &= \overline{R(0)-R(f^{\mathrm{st}})}, \quad G^{\mathrm{sw}} = \overline{R(0)-R(f^{\mathrm{sw}})}, \\ D_{\mathrm{AR}} &= \overline{R(f^{\mathrm{sw}})-R(f^{\mathrm{st}})-R(f^{\mathrm{sw}}_{\mathrm{AR}})+R(f^{\mathrm{st}}_{\mathrm{AR}})}, \\ W &= \min(G^{\mathrm{st}}, G^{\mathrm{sw}}). \end{aligned} \tag{3}$$

The overbar averages the three primary offsets. A positive $G$ means lower conditional expected error than zero; we call this a useful forecast. Positive $W$ requires usefulness in both histories on the same path. A negative $D$AR means that the model's increase in error between stable and switched histories is smaller than the increase for full-context AR. This last comparison measures relative disruption, which need not imply useful prediction.

Each process is evaluated on 1,024 independently seeded paths. Returns within a path can be dependent, and forecasts from different models, histories and offsets on that path are paired observations. The independent sample size is therefore 1,024. We use simultaneous distributional bounds based on DKW–Massart (Massart, 1990) and direct paired probability contrasts based on empirical Bernstein bounds (Maurer & Pontil, 2009). Together they cover the reported probability statements across the three processes with at least 95% unconditional joint coverage under the stagewise sampling design. They do not provide coverage conditional on outcomes already inspected. This secondary analysis allocates .025/3 to each of the distributional and paired-contrast families within a process; the original .05/3 distributional allocation is a separate analysis.

Probability alone does not describe the size of a forecast gain. We therefore also examine gains capped symmetrically at $\pm c$, whose mean is $\mathrm{E}[\mathrm{clip}(G, -c, c)]$. Integrating the simultaneous CDF bounds gives mean bounds that hold for every positive cap on the same event. This provides inference for a bounded quantity, not for the original untrimmed mean. We show a 30 bp cap in the main results. The untrimmed means, offset-specific results and reference frequencies remain descriptive.

## 3 What conflicting history changes

### 3.1 Stable signal recovery does not ensure resilience

The models' response to conflicting history is most informative where they can already extract a stable signal. At the higher benchmark, all five models improve on zero on a majority of paths under both stable dependence signs and in all three processes. Conflicting older observations reduce that probability in all 30 comparisons (Figure 3). In the financially calibrated process, stable usefulness ranges from 80.47% to 95.80%; even the smallest simultaneous lower bound is 73.53%. The deterioration is therefore not explained by an inability to learn the stable relationship.

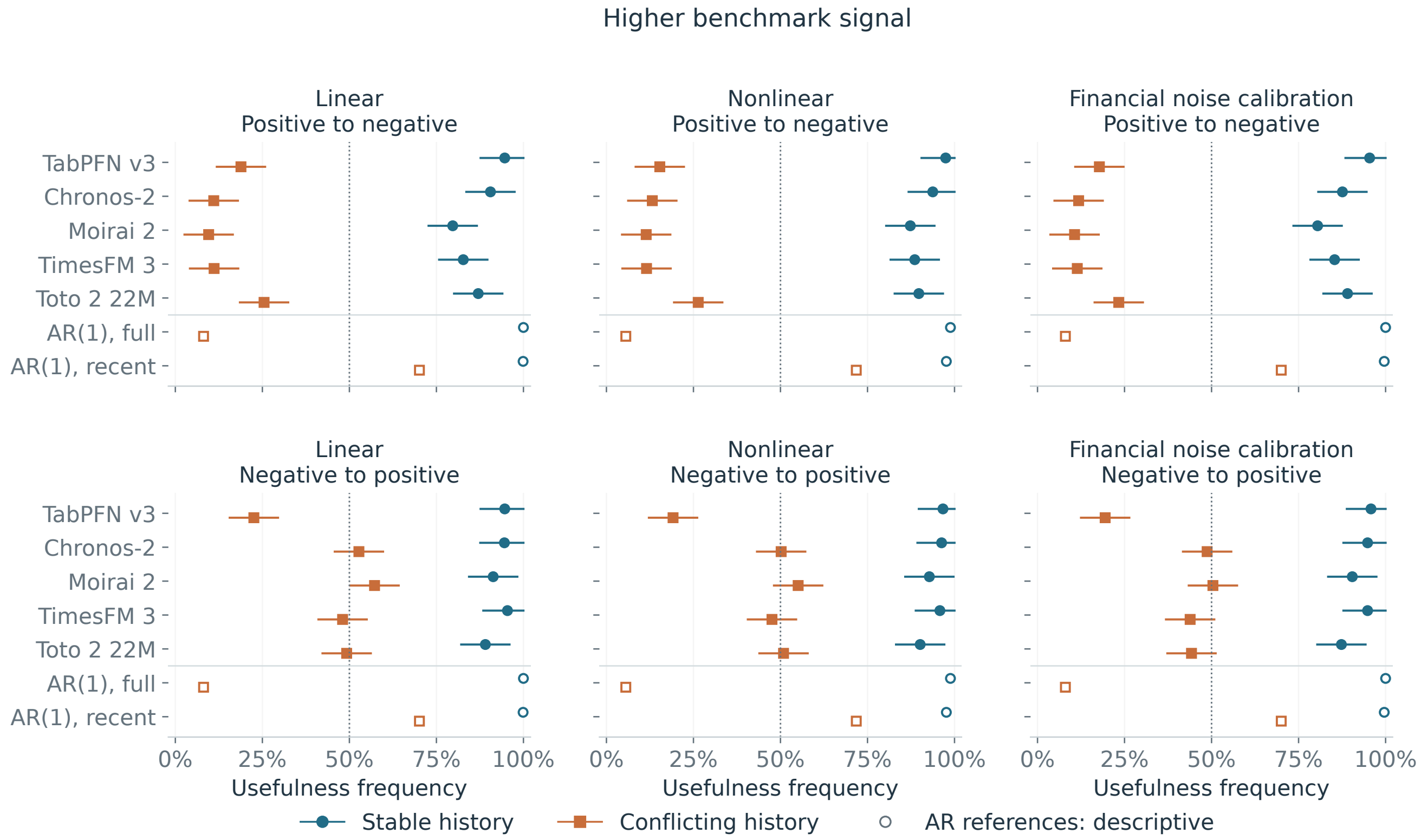

**Figure 3.** Probability of lower conditional error than zero at the higher benchmark signal, after averaging offsets 16, 64 and 128 within each path. Filled circles and squares show stable and switched histories with simultaneous CDF bounds. Hollow AR reference points are descriptive. The horizontal scale describes the frequency of positive gains, not their magnitude. Each panel uses the same 1,024 paths for all procedures.

At the lower benchmark, the upper usefulness bound for switched histories lies below one half in 29 of the 30 comparisons. The remaining comparison, Chronos under a negative-to-positive reversal in the nonlinear process, is unresolved. At neither signal level does any comparison establish a majority of paths on which the model is useful in both histories. These results concern the two signal levels studied; they do not imply monotone deterioration at smaller, untested signals.

### 3.2 Adjustment depends on the direction of change

The primary average conceals differences in how forecasts respond as the recent history grows. Figure 4 shows those differences in the higher-signal financial process. After a change from negative to positive dependence, Chronos and Moirai first exceed 50% empirical usefulness at the offset-64 checkpoint, reaching 64.26% and 65.04% at offset 128. After the opposite reversal, their offset-128 frequencies remain 43.75% and 42.29%. The destination relationship matters even though the paired construction preserves the available prediction problem within each comparison.

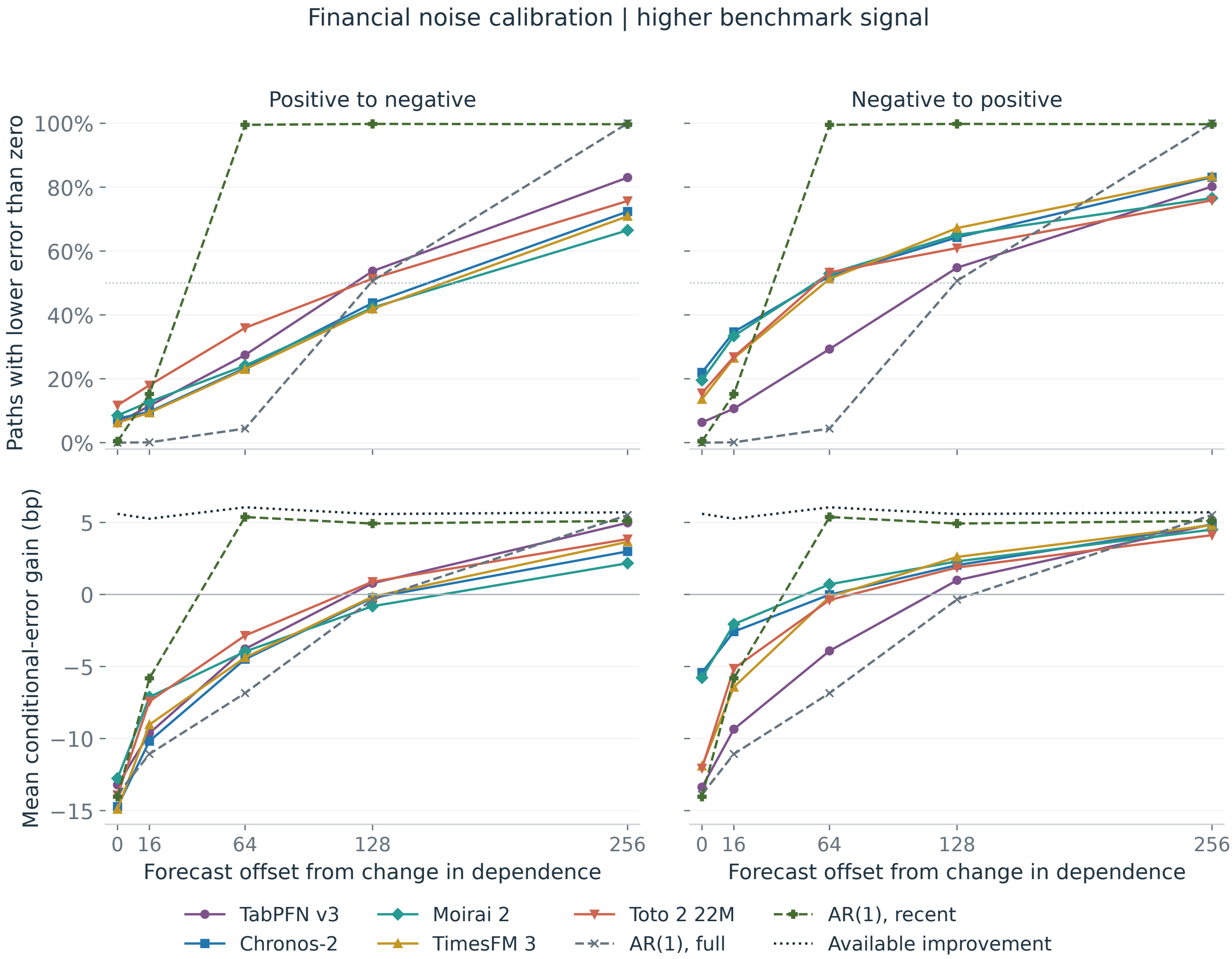

**Figure 4.** Higher-signal financial process. The upper panels show usefulness at each offset; the lower panels show descriptive untrimmed mean gains and the mean available improvement under the known law. Lines join five evaluated checkpoints and do not estimate intermediate responses. All curves use the same 1,024 paths. Stable and switched inputs coincide at offset 256.

Direct paired comparisons establish 16 positive directional contrasts out of 20 within each process, favoring negative-to-positive reversals. TabPFN's four directional contrasts remain unresolved. This is evidence of unequal responses for the resolved comparisons, rather than evidence that the unresolved model is invariant. Section 5 relates the pattern to a symmetry condition on the complete forecasting procedure.

Recent AR helps explain why timing matters. By offset 64, all of its fitted transitions follow the current relationship. In either reversal direction, empirical usefulness then reaches 99.51% and descriptive mean gain is 5.376 bp. At offset 16, however, its mean gain remains −5.816 bp. The favorable later response does not imply immediate recovery, and the five evaluated checkpoints do not identify an exact adjustment time.

### 3.3 Frequency, magnitude and relative disruption

The mean gains give a less favorable view of switched-history performance than stable usefulness alone might suggest. At the higher signal, all ten stable model means are positive in every process, but nine of the ten switched means are negative. In the financial process, Moirai after a negative-to-positive reversal is the sole model with a positive switched estimate, at 0.312 bp. Full-context AR has a switched mean of −6.088 bp, compared with 1.494 bp for recent AR. These untrimmed averages are descriptive; they do not come with a finite-sample mean guarantee.

| Procedure | Reversal | Stable mean (bp) | Switched mean (bp) | Capped switched mean [bounds] (bp) | Smaller AR penalty % [bounds] |
|---|---|---|---|---|---|
| TabPFN v3 | + to − | 4.861 | -4.210 | -4.155 [-6.555, -0.758] | 72.0 [65.0, 78.9] |
| TabPFN v3 | − to + | 4.833 | -4.092 | -4.042 [-6.457, -0.627] | 71.9 [64.9, 78.8] |
| Chronos-2 | + to − | 2.987 | -4.965 | -4.909 [-7.146, -1.457] | 81.4 [74.5, 88.4] |
| Chronos-2 | − to + | 4.768 | -0.186 | -0.186 [-2.834, 2.588] | 93.8 [86.9, 100.0] |
| Moirai 2 | + to − | 2.129 | -3.964 | -3.941 [-6.103, -0.817] | 90.8 [83.9, 97.8] |
| Moirai 2 | − to + | 4.386 | 0.312 | 0.312 [-2.426, 2.898] | 93.5 [86.5, 100.0] |
| TimesFM 3 | + to − | 3.798 | -4.520 | -4.431 [-6.720, -1.120] | 79.7 [72.7, 86.6] |
| TimesFM 3 | − to + | 4.791 | -1.327 | -1.307 [-4.011, 1.828] | 83.4 [76.5, 90.3] |
| Toto 2 22M | + to − | 3.849 | -3.129 | -2.729 [-5.161, 0.602] | 89.1 [82.1, 96.0] |
| Toto 2 22M | − to + | 3.728 | -1.218 | -0.964 [-3.705, 2.221] | 91.8 [84.9, 98.7] |

**Table 1.** Higher-signal financial process. Stable and switched means are untrimmed descriptive estimates. Capped switched means use ±30 bp and have simultaneous bounds. The last column reports the probability and simultaneous interval for a smaller history penalty than full-context AR. A basis point is 0.0001 of absolute return loss, not a trading return.

The capped means support a more limited statement about loss magnitude. At ±30 bp, five switched intervals lie below zero: TabPFN in both directions, and Chronos, Moirai and TimesFM after positive-to-negative reversals. The remaining five intervals include zero. Eight of the ten stable capped-mean intervals lie above zero. Increasing the cap changes both the estimand and its precision.

Relative disruption tells a different part of the story. All ten higher-signal financial comparisons establish a smaller history penalty than full-context AR on a majority of paths, with empirical frequencies from 71.88% to 93.85%. Yet most switched mean gains are negative. A model can therefore be less disrupted than this reference and still deliver an unfavorable average forecast gain. Robustness comparisons should be read alongside performance against zero.

## 4 Does a shorter history help on observed returns?

### 4.1 A matched comparison of input lengths

The simulation identifies an effect of older observations under a known law. The historical exercise asks whether a simple response—using fewer observations—improves the same procedures on real returns. The panel contains 56 selected assets and 434 forecast origins from January 2, 2025 through September 25, 2026, with final targets on September 28. For each of the 24,304 asset-origin pairs per model, the two input lengths predict the same next adjusted-close log return.

The shorter input contains exactly the last 64 returns of the 256-return input. Model checkpoints, preprocessing conventions and output medians are held fixed. For TabPFN, retaining eight lag features reduces the in-context training sample from 248 to 56 rows. Thus truncation changes the information and, in that case, the effective training sample available to the procedure. It does not label the removed observations as obsolete. Finite replay checks support reuse of the archived long-context forecasts without establishing invariance for arbitrary inputs or random streams.

The primary difference is MAE64 − MAE256, averaged over origins within each asset and then equally over assets. Negative values favor the shorter history. We use 5,000 circular bootstrap resamples of complete dates, keeping all assets, models and both lengths together. The primary block length is 20 origins, with sensitivities of 5 and 60. Table 2 reports pointwise percentile intervals and centered two-sided p-values with Holm adjustment across the five models (Holm, 1979). The panel and long-context outcomes were inspected before this extension was specified, so the comparison is exploratory and its bootstrap inference is approximate.

### 4.2 Shortening raises error for every model

All five models have higher estimated error with the shorter input (Table 2; Figure 5). The increase ranges from 1.769 bp for Toto to 4.581 bp for Chronos. Each primary difference meets the .05 Holm criterion, and both sensitivity block lengths retain positive intervals. The result is consistent across the tested models, although it concerns one selected panel and two input lengths rather than a universal advantage of longer histories.

| Procedure | MAE256 | MAE64 | Difference [95% interval] | Holm p | Gain256 | Gain64 |
|---|---|---|---|---|---|---|
| TabPFN v3 | 155.143 | 158.220 | 3.077 [2.297, 3.957] | 0.001 | -0.321 | -3.398 |
| Chronos-2 | 155.268 | 159.849 | 4.581 [3.734, 5.650] | 0.001 | -0.445 | -5.027 |
| Moirai 2 | 156.113 | 159.717 | 3.604 [2.824, 4.521] | 0.001 | -1.291 | -4.895 |
| TimesFM 3 | 156.850 | 160.575 | 3.725 [3.052, 4.597] | 0.001 | -2.028 | -5.753 |
| Toto 2 22M | 156.195 | 157.964 | 1.769 [1.316, 2.232] | 0.001 | -1.373 | -3.141 |

**Table 2.** Historical accuracy in basis points of absolute return loss. Difference is MAE64 − MAE256, so positive values favor 256. Gain is MAEzero − MAEmodel, so positive values favor the model; zero MAE is 154.822 bp. Intervals are pointwise approximate 95% date-block bootstrap intervals. Raw p-values are 1/5001 and Holm values are 5/5001 (rounded to .001), the fixed resampling resolution floor. Zero-relative gains are descriptive.

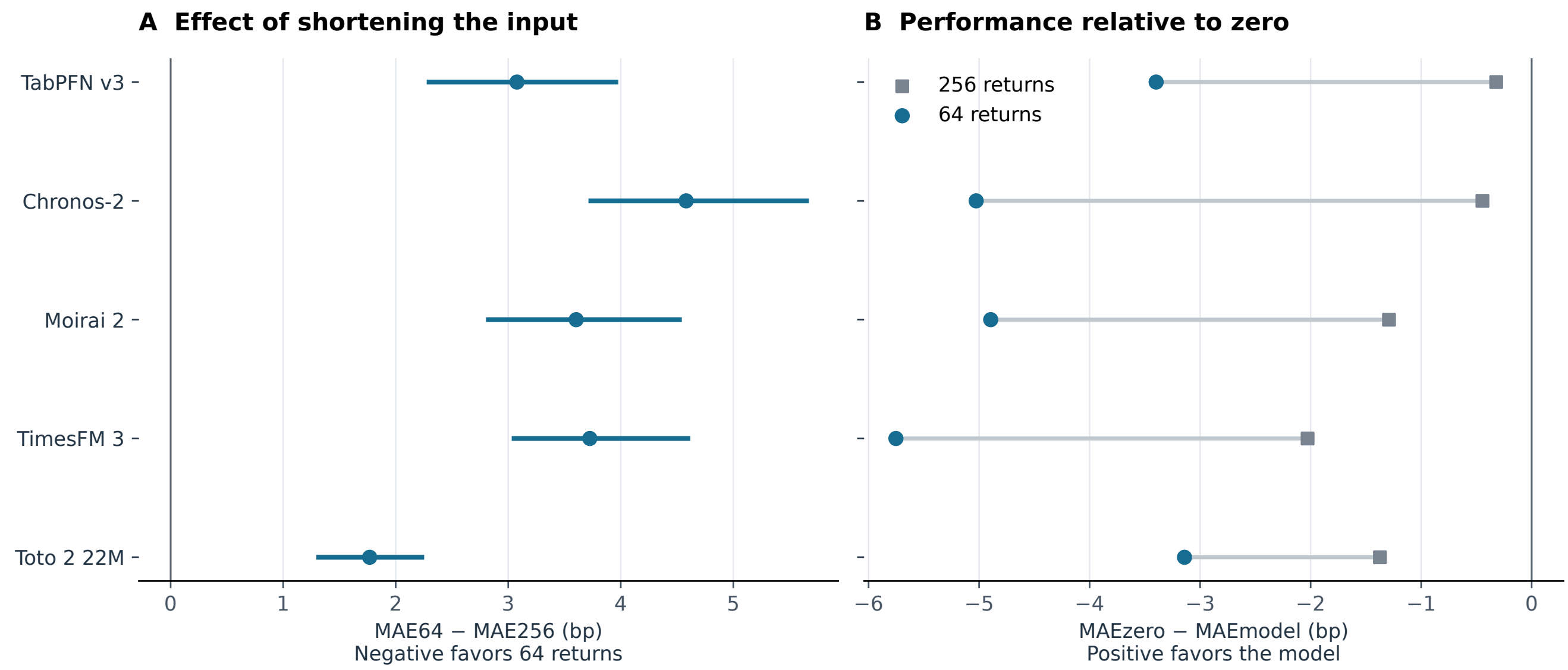

**Figure 5.** Panel A shows the paired equal-asset MAE difference with pointwise approximate 95% intervals from 5,000 circular date-block resamples of length 20. Positive differences favor 256 returns. Panel B shows descriptive gains relative to zero for each input length. The intervals are not simultaneous; Table 2 reports Holm adjustment across the five models. Units are absolute return-loss basis points.

Neither input length improves observed overall MAE over zero. The short-context gains relative to zero range from −3.141 to −5.753 bp, compared with −0.321 to −2.028 bp for the long context. This distinction is consequential: choosing the better of two input lengths does not by itself produce a useful return forecast.

The archived simple references provide a descriptive comparison on the same asset-origin panel. Full-context AR has an MAE of 155.318 bp, compared with 156.794 bp for recent AR, a difference of 1.476 bp. Recent AR fits 64 lag pairs from 65 returns; full-context AR fits 255 pairs from 256 returns. This window contrast differs from the 64-versus-256 comparison in Table 2. The 256-return historical-median reference has an MAE of 154.825 bp. These descriptive reference results are outside Table 2's five-model inferential family.

The same descriptive direction holds after large observed moves, in their complement, and in three volatility groups. All 25 subgroup model differences favor 256 returns. After large moves, shortening raises estimated error by 4.109–13.603 bp. These are descriptive comparisons without subgroup significance claims. Thresholds and eligibility rules are fixed from the development design.

One possible explanation is that fewer observations make the forecasting procedure more sensitive to recent shocks. The present comparison does not distinguish this mechanism from changes in normalization, representation or effective in-context sample size.

Taken together, the historical results qualify the practical implication of the simulation. Conflicting observations can impair a forecaster, but an unconditional rule that removes older returns also removes potentially useful information. Sensitivity to conflicting history is therefore insufficient evidence for adopting shorter inputs on this panel.

## 5 Interpretation and limits

The directional results admit a precise comparison. Let $\tau$ denote the index at which the dependence changes. Multiplying returns by the alternating sequence $A_t = (-1)^{t-(\tau-1)}$ exchanges the two dependence signs while preserving the conditional scale. Absolute risk satisfies $R(A_t f;\ A_t\mu,\ \sigma) = R(f;\ \mu,\ \sigma)$. A forecasting rule with the corresponding equivariance property,

$$f(A_{\text{history}} \odot H) = A_t f(H). \tag{4}$$

would therefore have equal risks and gains path by path under the transformed reversal directions. Here $\odot$ denotes elementwise multiplication of the history $H$. Unequal paired risks reveal a failure of this property on the evaluated inputs. Equal aggregate outcomes would not establish it. The criterion describes the complete forecasting procedure; it does not locate the source of asymmetry in pretraining, normalization, representation or the output rule. In particular, TabPFN's unresolved direction contrasts do not prove invariance or establish an explanation based on its lag representation.

The exact matching construction has a specific domain. It relies on symmetric innovations, an odd conditional-location function and volatility driven by squared shocks. Skewness and leverage effects fall outside it. The nonlinear calibration matches one average measure of predictability rather than its entire distribution, and the financial calibration conditions on a single fitted sample without propagating parameter uncertainty. Differences across processes do not isolate tail shape, scale or volatility persistence, which change together. Nor does the experiment establish a vulnerability unique to foundation models.

The statistical scope is equally important. The secondary paired-contrast and capped-mean analyses were specified after inspection of the initial distributional results, and the historical context-length comparison was specified after inspection of the long-context panel. Simultaneous bounds cover the stated bounded quantities under the specified sampling laws; they do not convert exploratory design choices into independent validation. In particular, capped-mean inference cannot resolve the contribution of unseen extreme losses to the untrimmed mean. Conclusions at the two benchmark signals do not extend to untested coefficients without further assumptions.

For the historical comparison, retrospective asset selection, overlapping inputs, dependence across dates, possible pretraining overlap, checkpoint chronology and adjusted-price vintage limit external validity. The bootstrap is an approximation based on 434 dates. Truncating an input is not equivalent to observing a change in the underlying market law, and neither the simulated gains nor the historical loss differences are portfolio returns. Context adaptation may be worth evaluating, but these results establish neither its necessity nor its sufficiency for useful return prediction.

## 6 Conclusion

The ability to recover a stable signal does not ensure that a forecasting model will remain useful when older observations reflect the opposite relationship. By matching recent returns and the conditional target distribution, the paired design isolates this distinction. Across five models and three processes, conflicting history reduces higher-signal usefulness in every comparison, with directional differences and loss magnitudes that matter beyond average accuracy.

The historical evidence cautions against turning that finding into a simple truncation rule. On the matched daily-return panel, 64-return inputs perform worse than 256-return inputs for every model, and neither length improves observed overall error over zero. The contribution is therefore an evaluation principle and a controlled design: historical sensitivity should be assessed together with signal recovery and the magnitude of forecast gains before it motivates a change in how models use the past.

Additional methods, mathematical derivations and supporting results are available from the author upon request.